\documentclass[12pt,thmsa]{article}
\usepackage{amssymb}

\input{tcilatex}
\begin{document}

\author{Karl-Georg Schlesinger \\
Institute for Theoretical Physics\\
University of Vienna\\
Boltzmanngasse 5\\
A-1090 Vienna, Austria\\
e-mail: kgschles@esi.ac.at}
\title{A physics perspective on geometric Langlands duality}
\date{}
\maketitle

\begin{abstract}
We review the approach to the geometric Langlands program for algebraic
curves via S-duality of an N=4 supersymmetric four dimensional gauge theory,
initiated by Kapustin and Witten in 2006. We sketch some of the central
further developments. Placing this four dimensional gauge theory into a six
dimensional framework, as advocated by Witten, holds the promise to lead to
a formulation which makes geometric Langlands duality a manifest symmetry
(like coavariance in differential geometry). Furthermore, it leads to an
approach toward geometric Langlands duality for algebraic surfaces,
reproducing and extending the recent results of Braverman and Finkelberg.
\end{abstract}

\section{Introduction}

In April 2006 Kapustin and Witten published their pathbreaking work \cite{KW}
which led to a completely new perspective on geometric Langlands duality for
algebraic curves. It starts from a four dimensional $N=4$ supersymmetric
gauge theory. Assuming that $S$-duality, a certain symmetry which
generalizes the electric-magnetic duality of the Maxwell equations to the
case of a nonabelian gauge theory, holds for this theory, it is possible to
derive geometric Langlands duality for algebraic curves from this. $S$%
-duality is conjectural but very well supported. In this sense, we get a
reformulation of geometric Langlands duality for algebraic curves.

In the first part of this contribution, we will review the approach of
Kapustin and Witten. We will continue by very briefly sketching some of the
new developments which this approach has initiated. Finally, we will place
the four dimensional gauge theory in a six dimensional string theory
framework. This perspective (see \cite{Wit 2007}, \cite{Wit 2009a}) holds
the promise to lead to a formulation, making geometric Langlands duality a
manifest symmetry (like covariance is manifest in differential geometry and
has no longer to be verified by calculations on specific coordinate
transformations). The six dimensional view also leads to an approach toward
geometric Langlands duality for algebraic surfaces, reproducing and
extending the recent results of Braverman and Finkelberg (see \cite{BF}, see
also \cite{Nak}).

This article is intended as an introduction to the gauge and string theory
approach to the geometric Langlands program for mathematicians. As such, it
focuses on a short, non-technical, overview of the central ideas and
concepts and does not contain any original research results. Neither do we
pretend to give a complete overview of this rapidly developing and highly
promising field. To keep the article in a sufficiently focused form, some
exciting developments (e.g. the appearance of Arthur's $SL\left( 2\right) $
in this framework, see \cite{BN} and \cite{Wit 2009b}) will completely be
left out. For another recently published review on the topic, see \cite{Fre2}%
.

I would like to thank A. Schmitt for the invitation to contribute this
article and to give a talk on a similar topic at the highly stimulating
conference VBAC 2009 in Berlin.

\bigskip

\section{N=4 supersymmetric gauge theory}

Let $G$ be a compact Lie group. For simplicity (and to keep all the formulae
valid in precisely the form used here, without any extra factors), we will
assume that $G$ is from the $ADE$-series. Later, for the six dimensional
framework, this assumption will be essential and no longer a technical
assumption for simplicity.

Let $X_4$ be the four dimensional space-time (again, we make a technical
assumption for simplicity, assuming that the signature of $X_4$ is Euclidean
rather than Minkowskian). We use Greek indices, e.g 
\[
\mu =0,\ 1,\ 2,\ 3 
\]
to label space-time indices and Latin ones 
\[
i=1,...,6 
\]
for an internal set of indices. Let $A$ be a connection of a $G$-bundle over 
$X_4$ and $F$ the corresponding curvature form. With 
\[
D_{\mu} 
\]
we denote the covariant derivatives and with 
\[
\phi _i 
\]
a set of adjoint-valued scalar fields (i.e. functions valued in the adjoint
representation of the Lie algebra of $G$). The action of the $N=4$
supersymmetric gauge theory which we want to consider is then given by 
\begin{eqnarray*}
S_4 &=&\frac 1{e^2}\int d^4x\ Tr\left( \frac 12\sum_{\mu ,\nu =0}^3F_{\mu
\nu }F^{\mu \nu }\ +\sum_{\mu =0}^3\sum_{i=1}^6D_{\mu} \phi _iD^{\mu} \phi
_i+\frac 12\sum_{i,j=1}^6\left[ \phi _i,\phi _j\right] ^2\right) \\
&&+\ ...
\end{eqnarray*}
where the dots indicate the fermionic part of the action and $e$ the
coupling constant of the theory (just as Newton's constant in gravity). The
fermionic part is necessary for supersymmetry but we will not consider it,
here. Though we can explain the essential ideas without considering it
explicitly, one should nevertheless keep in mind that the whole construction
does only hold with $N=4$ supersymmetry implemented. Here, $N=4$ denotes the
degree of supersymmetry. The supersymmetry algebra is determined by a choice
of representation of the -- in this case four dimensional -- spin group,
i.e. the double cover of the Lorentz group. The simplest degree of
supersymmetry is denoted by $N=1$ while $N=4$ means (very roughly speaking)
that we have four copies of the simplest representation involved in the
definition of the supersymmetry algebra.

If one has higher than $N=1$ supersymmetry, it is generally possible to
derive the theory from an $N=1$ supersymmetric theory, living in a higher
dimensional space, by dimensional reduction. In this case, the higher
dimensional theory is $N=1$ supersymmetric gauge theory in ten dimensions.
Dimensional reduction (which will appear over and over again in this
article)\ means that one assumes some of the dimensions (in this case six)\
to be scrolled up to a compact space of very small volume (in this case, six
dimensions are scrolled up to small circles). Sending the volume to zero
(i.e. sending the radii of the circles to zero) results in an induced lower
dimensional theory. The fact that we can get the action $S_4$ from ten
dimensions by reducing on six circles is the reason for the appearance of
the adjoint-valued scalar fields and the internal indices $i=1,...,6$.
Indeed, the first term in $S_4$ is the well known gauge theory term while
the other two arise from the dimensional reduction of the corresponding
gauge theory term in ten dimensions.

To get the most general $N=4$ supersymmetric gauge theory in four
dimensions, it is possible to add the so called \textit{topological term} $%
S_{\theta} $ to $S_4$. This is given by 
\[
S_{\theta} =-\frac \theta {8\pi ^2}\int d^4x\ Tr\left( F\wedge F\right) 
\]
It is referred to as topological since the integral just gives the second
Chern class of the $G$-bundle. The coupling constant $e$ of $S_4$ and the
parameter $\theta $ of the topological term are combined into the complex
coupling constant 
\[
\tau =\frac \theta {2\pi }+\frac{4\pi i}{e^2} 
\]
Observe that the imaginary part of $\tau $ is always positive, i.e. $\tau $
is from the upper half plane $\Bbb{H}$.

\bigskip

\section{S-duality}

In the quantized theory there is a natural symmetry given by the generator 
\[
T:\tau \mapsto \tau +1 
\]
which results from the fact that -- roughly speaking -- the complex coupling
constant appears only in the form 
\[
e^{2\pi i\tau } 
\]
in the path integral.. On the other hand, there is a natural action of $%
SL\left( 2,\Bbb{R}\right) $ on $\Bbb{H}$.

The $S$\textit{-duality conjecture} states the following:

\bigskip

There exists a second symmetry (i.e. generator) $S$, generating together
with $T$ a discrete subgroup of $SL\left( 2,\Bbb{R}\right) $ such that $%
S_4+S_{\theta} $ is invariant under the following combination of operations:

\begin{itemize}
\item  $\tau \mapsto S\left( \tau \right) $

\item  exchange of electric and magnetic charges

\item  $G\mapsto \ ^LG$
\end{itemize}

\bigskip

The latter two operations are not unrelated: In 1977 Goddard, Nuyts, and
Olive investigated how electric and magnetic charges are classified in a
nonabelian gauge theory (see \cite{GNO}). The result is that one set of
charges is given by the weight lattice of the Lie algebra of the gauge group 
$G$ while the other one is given by the root lattice. Of course, exchanging
weight and root lattice is precisely what defines the Langlands dual $^LG$
of the group $G$.

In the same year Montonen and Olive presented $S$-duality as a conjectural
symmetry for nonabelian gauge theory (see \cite{MO}). While $S$-duality does
not hold in the non-supersymmetric case, there is strong evidence that it
holds with $N=4$ supersymmetry.

For $G$ from the $ADE$-series (as we do assume), the generator $S$ has to
take the form 
\[
S:\tau \mapsto -\frac 1\tau 
\]
i.e. the discrete subgroup of $SL\left( 2,\Bbb{R}\right) $ generated by $T$
and $S$ is the modular group $SL\left( 2,\Bbb{Z}\right) $.

\bigskip

In order to derive the geometric Langlands duality for algebraic curves from
the $S$-duality conjecture, one has to perform two essential steps on the
four dimensional gauge theory: First, one has to perform a topological twist
and than a dimensional reduction to a two dimensional theory. We will
discuss both steps very briefly, in a non-technical manner, in the next two
sections. After that we will introduce the operators of the four dimensional
gauge theory which -- after performing the two steps on them -- will lead to
geometric Langlands duality. For technical details, we refer the reader to 
\cite{KW}.

\bigskip

\section{Topological twisting}

Topological twisting means that one retains only part of the state space of
the original theory. For this one introduces the cohomology with respect to
a certain differential $Q$ (what physicists call a BRST-operator) with 
\[
Q^2=0 
\]

To find a $Q$ suitable for the topological twist, supersymmetry is
essential. Passing to the cohomology with respect to $Q$ -- i.e. forgetting
all $Q$-exact terms -- one retains only part of the information of the
original theory. The resulting theory is called topological. For a pure
mathematician this nomenclature might be slightly disturbing since the
theory is \textit{not} independent of all non-topological information, e.g.
we will see that after dimensional reduction it is still dependent on
certain holomorphic and symplectic structures. Topological theory in this
context means that \textit{on} the $Q$-cohomology we have independence from
the choice of metric.

Concretely, in this case $Q$ is determined by a choice of homomorphism 
\[
\chi :Spin\left( 4\right) \rightarrow Spin\left( 6\right) 
\]
which is related to the fact that the gauge theory arises from a ten
dimensional theory by dimensional reduction and a decomposition of $%
Spin\left( 10\right) $ into $Spin\left( 4\right) $ and $Spin\left( 6\right) $
components. The approach is very similar to the introduction of Donaldson
invariants for four dimensional manifolds by using a topological twist for $%
N=2$ supersymmetric four dimensional gauge theory.

It turns out that the topological twist is not determined uniquely but there
arises a whole family of suitable topological twists, parametrized by the 
\textit{topological twisting parameter} 
\[
t\in \Bbb{C}P^1 
\]
The complex coupling constant $\tau $ and the topological twisting parameter 
$t$ are then combined into the \textit{canonical parameter} 
\[
\psi =\frac{\tau +\overline{\tau }}2+\frac{\tau -\overline{\tau }}2\left( 
\frac{t-t^{-1}}{t+t^{-1}}\right) 
\]
The reason for introducing the canonical parameter is that the correlation
functions of the observables of the topological theory (i.e. of the $Q$%
-cohomology classes) do only depend on $\psi $ and not on the parameters $e$%
, $\theta $, and $t$ separately.

It is a small lemma to show that on $\psi $ the two generators $T$ and $S$
operate, again, as 
\[
T:\psi \mapsto \psi +1 
\]
and 
\[
S:\psi \mapsto -\frac 1\psi 
\]

\bigskip

\section{Dimensional reduction}

Let now 
\[
X_4\cong \Sigma \times C 
\]
with $\Sigma $ a (compact or non-compact) Riemann-surface (which will become
the two dimensional space-time after dimensional reduction) and $C$ a
compact Riemann surface. As a technical assumption, we will require that 
\[
genus\left( C\right) \geq 2 
\]
We now perform the dimensional reduction by assuming that 
\[
vol\left( C\right) \ll vol\left( \Sigma \right) 
\]
In this limit, the four dimensional $N=4$ supersymmetric gauge theory
induces a two dimensional field theory on $\Sigma $. In two dimensions there
do not exist non-trivial gauge theories and the resulting field theory turns
out to be a nonlinear sigma model, i.e. a field theory where the (bosonic)\
fields are given by maps from $\Sigma $ to the so called \textit{target space%
}. Roughly speaking, the action is given by a minimal area requirement for
the image of $\Sigma $ under these maps into the target space. So, the
essential information to determine the nonlinear sigma model is to specify
the target space. One shows that in this case the resulting target space is
the Hitchin moduli space $Hit\left( G,C\right) $ for the gauge group $G$ and
the complex curve $C$ (see \cite{Hit}).

\bigskip

Let $A$ be a $G$-connection on $C\,$and $F$ the corresponding curvature
form. Let $\phi $ be an adjoint-valued 1-form on $C$. Consider the set of
equations 
\[
F-\phi \wedge \phi =0 
\]
and 
\[
D\phi =D^{*}\phi =0 
\]
The solutions to this set of equations, modulo $G$-gauge transformations,
define the Hitchin moduli space $Hit\left( G,C\right) $. For those readers
with a knowledge of Higgs-bundles one can simply define it as the moduli
space of $G$-Higgs-bundles on $C$. That we have used the letter $\phi $ for
the adjoint-valued 1-form is not by accident. The topological twist shifts
the degree of some fields and we get the adjoint-valued 1-form from the
adjoint-valued scalar fields of the original gauge theory. With 
\[
g=genus\left( C\right) 
\]
one has 
\[
\dim _{\Bbb{C}}Hit\left( G,C\right) =\left( 2g-2\right) \dim G 
\]
and $Hit\left( G,C\right) $ is a Hyperk\"{a}hler manifold. So, we have a
representation of the quaternion algebra on the tangent bundle and complex
structures $I$, $J$, $K$ with corresponding symplectic structures $\omega _I$%
, $\omega _J$,$\omega _K$. When we refer to complex or symplectic structures
in the sequel, we will have to keep in mind that we have to make precise to
which of these structures we refer.

\bigskip

\section{Wilson operators}

We are now ready to introduce the needed operators in the four dimensional
gauge theory. Usual operators in a quantum field theory, as you remember
them from any introductory course on the subject, are attached to points
(i.e. they are zero dimensional objects): They are operators $M\left(
x\right) $, $M\left( z\right) $ attached to points $x$,$z$ and satisfying
the well known commutation relations (e.g. $M\left( x\right) $ and $M\left(
z\right) $ commute if $x$ and $z$ are space-like separated). Physicists have
learned in recent decades that there are other operators, attached to lines
(one dimensional objects), containing essential information in a quantum
field theory (in solid state physics or in the study of phase transitions
these are prominent operators).

Recall that $A$ is a connection on a $G$-bundle over $X_4$. Let $S$ be an
oriented loop in $X_4$, $R$ an irreducible representation of $G$. With $Tr_R$
we denote taking the trace in the representation $R$. We define the \textit{%
Wilson operator} $W_0\left( R,S\right) $ as the holonomy of $A$ around $S$: 
\[
W_0\left( R,S\right) =Tr_R\exp \left( -\oint_SA\right) 
\]
Since we want to perform the two steps, topological twisting and dimensional
reduction, on the operators, the next question is if these operators induce
well defined operators on $Q$-cohomology. Unfortunately, the answer is no
and this problem can not be resolved for general values of the topological
twisting parameter $t$. But for the special values $t=i$ and $t=-i$ there
exists a solution:\ For these values there exists a linear combination of $A$
with the adjoint-valued 1-form $\phi $, such that the holonomy of the linear
combination induces a well defined operator on cohomology, i.e. we have 
\textit{topological Wilson operators}. Concretely, the topological Wilson
operators are defined by 
\[
W\left( R,S\right) =Tr_R\exp \left( -\oint_SA+i\phi \right) 
\]
for $t=i$ and 
\[
W\left( R,S\right) =Tr_R\exp \left( -\oint_SA-i\phi \right) 
\]
for $t=-i$.

Next, replace the loop $S$ with a line $L$ from $p$ to $q$. Replace the
trace $Tr_R$ with the matrix of parallel transport from the fiber $E_p$ of
the $G$-bundle on $X_4$ to the fiber $E_q$, with both fibers considered in
the representation $R$ of $G$. The parallel transport is taken with respect
to the connection 
\[
\mathcal{A}=A+i\phi 
\]
and 
\[
\overline{\mathcal{A}}=A-i\phi 
\]
for $t=\pm i$, respectively. This corresponds to the canonical parameter $%
\psi =\infty $. In conclusion, for $\psi =\infty $ we have topological
Wilson operators, defined by representations $R$ of $G$.

Assume, now, that $S$-duality holds for the $N=4$ supersymmetric gauge
theory. This means that there has to exist a second set of topological line
operators which exchange with the Wilson operators on lines under $S$%
-duality, i.e. for $\psi =0$ there should exist topological line operators,
defined by representations of $^LG$. Indeed, these operators can be
constructed in the form of the so called \textit{'t Hooft operators} which
we are not going to discuss explicitly, here.

\bigskip

Finally, we perform the dimensional reduction on the topological line
operators. Consider a two dimensional theory and a line operator $\widehat{L}
$ on a line $L$, close to a boundary with specified boundary condition (i.e.
what physicists call a $D$-brane for a two dimensional nonlinear sigma
model). Imagine $L$ approaching the boundary more and more closely. In the
limit, $L$ will be absorbed by the boundary and the operator $\widehat{L}$
disappears, resulting in a change of boundary conditions. Of course, this is
a heuristic picture but it can be validated in a calculation. The boundary
condition is given by a submanifold ($D$-brane) of the target space to which
the one dimensional boundary of $\Sigma $ has to be mapped under the fields,
together with a vector bundle $W$ on this submanifold. One can show the
operator $\widehat{L}$ to change boundary conditions by changing this vector
bundle $W$. So, we can view the line operators in the two dimensional theory
as abstract operators, operating on boundary conditions.

We call a boundary condition, given by $W$, an \textit{eigenbrane} of $%
\widehat{L}$ if there exists a fixed vector space $V$ such that $\widehat{L}$
acts as 
\[
\widehat{L}:W\mapsto V\otimes W 
\]
This is similar to eigenfunctions for operators in quantum mechanics, with
the function replaced by a vector bundle and the eigenvalue replaced by the
fixed vector space $V$. As in quantum mechanics, we can pose the question if
line operators $\widehat{L_1}$ and $\widehat{L_2}$ on lines $L_1$ and $L_2$
can have simultaneous eigenbranes. The answer is that they have simultaneous
eigenbranes iff 
\[
\left[ \widehat{L_1},\widehat{L_2}\right] =0 
\]
For the dimensional reduction of the topologically twisted $N=4$
supersymmetric gauge theory, one can show that there exist simultaneous
eigenbranes of \textit{all} topological Wilson operators. These eigenbranes
are called \textit{electric eigenbranes}. Similarly, there exist
simultaneous eigenbranes of \textit{all} topological 't Hooft operators and
these are called \textit{magnetic eigenbranes}.

\bigskip

\section{Mirror symmetry}

Without defining the three complex structures $I$, $J$, $K$ (and
corresponding symplectic structures) explicitly for $Hit\left( G,C\right) $
(see \cite{KW}), we recall that we have to keep them apart when referring to
a complex or a symplectic structure. For a nonlinear sigma model on a Ricci
flat K\"{a}hler manifold there exist to types of topological twists, called
the $A$- and the $B$-model (see \cite{Wit 1991}). The $A$-model couples only
to the symplectic structure of the target space and the $B$-model only to
the holomorphic structure. One proves that

\smallskip

\textit{Electric eigenbranes are elements of the bounded derived category of
coherent sheaves in complex structure J on }$Hit\left( G,C\right) $.

\smallskip

The elements of the bounded derived category of coherent sheaves are the $D$%
-branes for the $B$-model, referred to as $B$-branes in the physics
literature. Including reference to the complex structure $J$, they are
called $J_B$-branes. Similarly, one shows that

\smallskip

\textit{Magnetic eigenbranes are elements of the Fukaya category in
symplectic structure }$\omega _K$.

\smallskip

In physics terminology, this means magnetic eigenbranes are $K_A$-branes.
Mirror symmetry exchanges the $A$- and the $B$-model. One mathematically
rigorous formulation of mirror symmetry, called \textit{homological mirror
symmetry} (see \cite{Kon 1994}) states that Calabi-Yau manifolds $X\,$and $Y$
form a mirror pair if there is a suitable equivalence between the Fukaya
category of $X$ and the bounded derived category of coherent sheaves of $Y$
and vice versa (to make this technically precise, one does not really work
with simple categories but with a triangulated version of $A_\infty $%
-categories). $S$-duality of the four dimensional gauge theory induces
homological mirror symmetry for the $Hit\left( G,C\right) $ sigma model on $%
\Sigma $ or in more physics oriented language, $S$-duality induces mirror
symmetry between the $B$-model on $Hit\left( G,C\right) _J$ (corresponding
to $\psi =\infty $) and the $A$-model on $Hit\left( ^LG,C\right) _K$
(corresponding to $\psi =0$). Here, the subscripts refer to the complex,
respectively symplectic structure, used on Hitchin moduli space.

\bigskip

In the geometric Langlands program for algebraic curves $C$ one considers
two different moduli spaces: The moduli space $\mathcal{M}$ of flat $^LG_{%
\Bbb{C}}$-bundles on $C$ and the moduli space $\widetilde{\mathcal{M}}$ of
holomorphic $G$-bundles on $C$. On $\mathcal{M}$ one considers sheaves with
support at a point of $\mathcal{M}$ (skyscraper sheaves) and on $\widetilde{%
\mathcal{M}}$ one considers the so called Hecke eigensheaves. It is a
central part of \cite{KW} to show that the skyscrapers are in one-to-one
correspondence to the electric eigenbranes and the Hecke eigensheaves to the
magnetic eigenbranes. In consequence, if $S$-duality holds for the four
dimensional $N=4$ supersymmetric gauge theory, one can derive geometric
Langlands duality for algebraic curves.

At this point, the attentive reader might ask why one needs $S$-duality of
the four dimensional gauge theory for this and why one does not start
directly from the homological mirror symmetry conjecture for the $Hit\left(
G,C\right) $ sigma model. The answer is that mathematicians no very well
that $\widetilde{\mathcal{M}}$ is not a true moduli space (and can not be
for geometric Langlands duality to hold true) but a stack. The nonlinear
sigma model treats the target space in a first approach as a proper space.
If one takes the stacky nature into account, one rediscovers that one
actually derived the model from the four dimensional gauge theory, i.e. the
four dimensional viewpoint is essential for the geometric Langlands program
(see \cite{KW}).

There are further examples for the deep interplay between the structures,
naturally emerging from physics, and those needed for the mathematics of the
geometric Langlands program, in this approach. E.g. the Fukaya category as
it is originally defined (see \cite{Fuk}, see \cite{FOOO} for an approach to
a rigorous treatment), involves the Lagrangian submanifolds of $Hit\left(
G,C\right) $. But there exist additional $A$-branes on $Hit\left( G,C\right) 
$ which are only coisotropic submanifolds. A special such $A$-brane (called
the canonical coisotropic brane or c.c. brane, for short), corresponding to
a coisotropic submanifold of full dimensionality (i.e. isomorphic to $%
Hit\left( G,C\right) $ itself) and of rank one (i.e. the vector bundle $W$
on the brane is a line bundle) is used in \cite{KW} to show that the
magnetic eigenbranes satisfy the $D$-module property which is so important
for the Hecke eigensheaves in the geometric Langlands program.

Finally, there exists another physics motivated approach to the geometric
Langlands program for algebraic curves, using two dimensional conformal
field theory on $C$ to construct Hecke eigensheaves (see \cite{Fre} for a
beautiful review and the original literature). One might wonder how the two
approaches are related, one leading to a two dimensional nonlinear sigma
model on $\Sigma $, the other to a conformal field theory on $C$. It would
be possible to derive the conformal field theory approach on $C$ also from
the four dimensional gauge theory if one could prove the following property
to hold: The dual brane (under $S$-duality, respectively mirror symmetry) of
the c.c. brane should be a brane which has support on the space of opers of 
\cite{BD} (\cite{Wit}). The dual of the c.c. brane is a coisotropic brane of
rank \TEXTsymbol{>} 1 and is calculated in the gauge theory setting in \cite
{GaW 2008b}.

After this review of some of the central parts of \cite{KW}, we are now
ready to take a brief look on some of the further developments in 2006-2009.

\bigskip

\section{Higher dimensional operators}

As we have seen, beyond the usual zero dimensional operators (attached to
points) there are one dimensional line operators in a quantum field theory,
containing fundamental information. One might ask if there are further even
higher dimensional operators.. In a four dimensional theory, these could be
two dimensional (attached to surfaces) or three dimensional (attached to
volumes). Four dimensional operators would be trivial.

Two dimensional operators become important if the gauge connection $A$ has
singularities. So far, we have assumed $A$ to be holomorphic but one can
allow $A$ to be meromorphic, only, and to have singularities along surfaces.
The approach of \cite{KW} can be extended to this case and surface operators
take a central place, then. When Beilinson and Drinfeld developed the
geometric Langlands program, it was intended as an analogue to the classical
Langlands program, to get insides from a situation with additional
smoothness property. The case of a meromorphic gauge connection $A$
corresponds to what is called ramification in the classical Langlands
program. If $A$ has only simple poles, one has tame ramification, otherwise
one has the case of wild ramification (see \cite{GW 2006}, \cite{GW 2008}, 
\cite{Wit 2007a}; see \cite{Fre} for a discussion how structures in the
classical and the geometric Langlands program are analogous). Especially,
understanding wild ramification in the geometric case is believed to be
important for comparison to the classical Langlands program.

\bigskip

Three dimensional operators live on volumes and therefore divide the four
dimensional manifold $X_4$ into two halves.. They correspond to what
physicists call \textit{domain walls} in a gauge theory. Domain walls allow
to change the gauge group. On the one side, we have the gauge theory with
gauge group $G$ and on the other side the theory with gauge group $%
\widetilde{G}$. On the domain wall we have the three dimensional operator,
corresponding to specifying a boundary condition which ensures that the two
gauge theories join consistently along the domain wall.

In the classical Langlands program, beyond Langlands duality, changing the
group $G$ is a central ingredient, giving rise to Langlands functoriality.
It was an open question -- again of tremendous importance for comparing the
geometric to the classical case -- what constitutes the counterpart of
Langlands functoriality in the geometric Langlands program. Domain walls
lead to geometric Langlands functoriality. This is a subject very much in
its beginning. From the gauge theory side one has to get knowledge on the
three dimensional boundary conditions which involve data given in the form
of three dimensional quantum field theories (see \cite{FW}, \cite{GaW 2008a}%
, \cite{GaW 2008b}, \cite{Wit 2009a}).

In conclusion, higher dimensional operators on surfaces and volumes have
turned out to be very important for studying analogues of structures which
are central for the classical Langlands program.

\bigskip

\section{The six dimensional view}

Remember that our four dimensional gauge theory lives on $X_4$. There is a
conjecture, arising from string theory, which states that there exists a
conformally invariant field theory on 
\[
X_4\times T^2 
\]
such that in the limit of small $T^2$ (dimensional reduction), it induces
precisely the $N=4$ supersymmetric gauge theory on $X_4$.

On $T^2$ there is, of course, the natural action of $SL\left( 2,\Bbb{Z}%
\right) $. We can ask what compensates this action on $X_4$. It turns out
that in this way the $SL\left( 2,\Bbb{Z}\right) $-action on $T^2$ induces $S$%
-duality of the gauge theory on $X_4$. In consequence, if it would be
possible to construct this six dimensional conformal field theory, one could
prove $S$-duality for the $N=4$ supersymmetric gauge theory on $X_4$ and, in
consequence, geometric Langlands duality. One should stress that existence
of the theory suffices: While for the four dimensional gauge theory one has
to prove something ($S$-duality) to get geometric Langlands duality, for the
six dimensional conformal field theory one only has to construct the theory
since its very existence makes $S$-duality of the four dimensional theory
(and, hence, geometric Langlands duality) manifest. In this sense, one can
view the search for this six dimensional theory as the search for the
geometry behind the Langlands program, making Langlands duality a manifest
symmetry. This would be very much like passing from a coordinate description
to differential geometry where covariance becomes a manifest symmetry.

The problem is that it is known from string theory that this six dimensional
theory can not exist consistently on its own. It actually has to be embedded
into eleven dimensional $M$-theory as the world-volume theory of the $M5$%
-brane (a five dimensional extended object with a six dimensional world
volume in $M$-theory, the central charges, leading to the $M5$-brane,
arising as one of the components in the direct sum decomposition of the
eleven dimensional supersymmetry algebra). So, its completion in the
UV-regime is related to the so called six dimensional micro string theories
(see \cite{Dij2}, \cite{DVV} for an easily accessible introduction).

\bigskip

Consider the six dimensional theory on another manifold $X_6$, now, 
\[
X_6\cong \Sigma \times X_4 
\]
with $\Sigma $ a (compact or non-compact) Riemann surface, $X_4$ a compact
Hyperk\"{a}hler manifold, and 
\[
vol\left( X_4\right) \ll vol\left( \Sigma \right) 
\]
This is very similar to the situation we considered when reducing the four
dimensional gauge theory to a two dimensional nonlinear sigma model. This
time we get the reduction of the six dimensional theory to a two dimensional
nonlinear sigma model and the target space turns out to be given by the
instanton moduli space $Inst\left( X_4\right) $ on $X_4$, i.e. the space of
all anti-self-dual $G$-connections on $X_4$ (remember that now, for the six
dimensional view, $G$ definitely has to be from the $ADE$-series).

From the side of physics, there are some very nice relations behind this
model. The Hitchin-Kobayashi correspondence relates $Inst\left( X_4\right) $
to $Bun_G\left( X_4\right) $, i.e. brings in a relation to Yang-Mills theory
on $X_4$. On the other hand, for 
\[
G=U\left( k\right) 
\]
the space $X_4$ turns out to be related to the multi-center Taub-NUT\
solution $TN_k$ of the Einstein vacuum equations. This gives particular
interest to the study of instantons on $TN_k$ (see \cite{Che 2008}, \cite
{Che 2009}, \cite{Wit 2009}).

On the mathematical side, this model reproduces and extends -- beyond the
case $G=U\left( k\right) $ the results of Braverman and Finkelberg (see \cite
{BF}, \cite{Nak}) on geometric Langlands duality for algebraic surfaces $X_4$
(see \cite{Tan}, \cite{Wit 2009a}). Let us review this in a little bit more
detail (basically following \cite{Wit 2009a}).

\smallskip

One can show that the six dimensional theory can not have a Lagrangian
description, it is a purely quantum field theoretic object. But
dimensionally reducing the theory for small $S^1$ on 
\[
X_6\cong X_5\times S^1 
\]
one gets in the infrared limit a gauge theory description on $X_5$. One can
now pass to the more complicated case with $X_6$ not being given as a
Cartesian product, as above, but as a $U\left( 1\right) $-bundle over $X_5$,
i.e. we have a free action of $U\left( 1\right) $ on $X_6$. This leads to an
additional Chern-Simons like term in the dimensional reduction to $X_5$.
Finally, one can pass to the case of a non-free action of $U\left( 1\right) $
on $X_6$ and consider the singular quotient space $X_6\diagup U\left(
1\right) $. Outside the non-free locus the dimensional reduction works as in
the previous case. Consider the special case where the non-free locus has
codimension four and consists only of fixed points of $U\left( 1\right) $.
In this case, the Chern-Simons term has an anomaly on the two dimensional
non-free locus $W$, i.e. on $W$ a third term has to appear in the action of
the dimensional reduction to $X_5$ which cancels this anomaly. This third
term arises from a two dimensional quantum field theory on $W$, given by the
holomorphic part of the $WZW$-model (at level one and for the group $G$).
The affine Lie algebra of $G$, which mathematically is behind the $WZW$%
-model, naturally explains why the approach to the geometric Langlands
program for algebraic surfaces (see \cite{BF}, \cite{Nak}) leads to
Langlands duality for the affine case.

\bigskip

\section{Conclusion}

We have seen that the search for a six dimensional field theory (which has
to be a purely quantum field theoretic structure, embedded into eleven
dimensional $M$-theory) offers a fundamental perspective on the geometric
Langlands program:\ It would lead to Langlands duality as a manifest
symmetry, it would unite geometric Langlands duality for algebraic surfaces
and algebraic curves into a single framework, and it would naturally include
higher dimensional operators which are so important for studying the
counterparts of ramification and Langlands functoriality on the geometric
side. In physics it has strong links to many areas (string- and $M$-theory,
Yang-Mills theory, Taub-NUT solutions of the Einstein equations).

Last not least, though the full six dimensional theory has not been
constructed so far, it is amenable to explicit calculations in dimensional
reductions, leading to structures like $WZW$-models where a lot of results
are available from the side of mathematical physics.

\end{document}